 \documentclass[3p,times]{elsarticle}
\usepackage{amssymb,amsmath}
\usepackage{float}
\usepackage{multirow}
\usepackage{hyperref}
\usepackage[figuresright]{rotating}

\newcommand{\LLc}{{\Lambda_c^+ \bar\Lambda_c^-}}
\newcommand{\eeLLc}{{e^+e^-\to \Lambda_c^+ \bar\Lambda_c^-}}
\newcommand{\ee}{{e^+e^-}}
\newcommand{\lbarl}{{\Lambda \bar\Lambda}}

\newcommand{\beq}{\begin{equation}}
\newcommand{\eeq}{\end{equation}}
\newcommand{\beqa}{\begin{eqnarray}}
\newcommand{\eeqa}{\end{eqnarray}}

\newcommand{\no}{\nonumber}

\def\no{\nonumber}
\def\bea{\arraycolsep .1em \begin{eqnarray}}
\def\eea{\end{eqnarray}}

\newlength{\feynwidth} 
\newlength{\feynwidthbig} 

\begin{document}

\begin{frontmatter}

\title{Re-examining the $X(4630)$ resonance in the reaction
$e^+e^-\rightarrow \Lambda^+_c\bar\Lambda^-_c$}

\author{Ling-Yun Dai$^1$}
\ead{l.dai@fz-juelich.de}
\author{Johann Haidenbauer$^1$}
\ead{j.haidenbauer@fz-juelich.de}
\author{Ulf-G. Mei{\ss}ner$^{2,1}$}
\ead{meissner@hiskp.uni-bonn.de}
\address{$^1$ Institut f\"ur Kernphysik, Institute for Advanced Simulation and
J\"ulich Center for Hadron Physics, Forschungszentrum J{\"u}lich, D-52425 J{\"u}lich, Germany}
\address{$^2$ Helmholtz Institut f\"ur Strahlen- und Kernphysik and Bethe Center
 for Theoretical Physics, Universit\"at Bonn, D-53115 Bonn, Germany}

\begin{abstract}
The reaction $e^+e^-\rightarrow \Lambda^+_c\bar\Lambda^-_c$ is investigated at energies
close to the threshold with emphasis on the role played by the $X(4630)$ resonance.
The interaction in the final $\Lambda^+_c \bar\Lambda^-_c$ system,
constructed within chiral effective field theory and supplemented by a pole diagram
that represents a bare $X(4630)$ resonance, is taken into account rigorously.
The pole parameters of the $X(4630)$ are extracted and found to be compatible
with the ones of the $X(4660)$ resonance that have been established in the reaction
$e^+e^- \to \pi^+\pi^-\psi(2S)$. The actual result for the $X(4630)$ is
$M = (4652.5\pm 3.4)$~MeV and $\Gamma = (62.6\pm 5.6)$~MeV.
Predictions for the $\Lambda^+_c$ electromagnetic form factors in the timelike region
are presented.
\end{abstract}

\end{frontmatter}

\section{Introduction}
\label{sec:1}

Over the last decade or so overwhelming experimental evidence has accumulated that casts some
doubts on our understanding of the hadron spectrum so far.  Specifically, at energies above the
open charm production threshold a plethora of structures were seen in experiments which do not
really fit into the standard picture that mesons are composed out of
quark-antiquark pairs. For recent overviews and discussions of these structures, commonly
referred to as X, Y and Z states, see for example \cite{Guo:2017,Chen:2016,Esposito:2016}.

Among these structures is a state listed as X(4660) in the latest compilation
of the Particle Data Group (PDG) \cite{PDG2016}. This X(4660) (also known as Y(4660))
was seen in the reaction $e^+e^- \to \pi^+\pi^-\psi(2S)$ \cite{Wang:2007,Lees:2014,Wang:2015}.
Additionally, a structure called the X(4630) was seen in the reaction $\eeLLc$ \cite{Belle08}
in a very nearby energy region. Finally, there is also an enhancement around $4660$ MeV
in the $\LLc$ invariant mass measured in the reaction $\bar B \to \LLc \bar K$ \cite{Aubert:2008}.
Since the mass and width derived from a Breit-Wigner based fit to the $\eeLLc$ data
yielded results that are consistent with those deduced from the $\pi^+\pi^-\psi(2S)$
channel it was already conjectured in Ref.~\cite{Belle08} that the states in
question could be the same.
The subsequent works by Bugg~\cite{Bugg:2009}, Cotugno et al.~\cite{Cotugno:2010},
and Guo et al.~\cite{Guo:2010} took up this interpretation and tried to corroborate it
with arguments and also with explicit calculations. Indeed, the PDG adopted likewise this point
of view by listing the states under the same heading \cite{PDG2016}. Note, however, that the
statement ``the states are not necessarily the same'' is added.
An entirely different issue is the dynamical origin of the state(s). While some
studies assign the X(4660) to a regular $c\bar c$ charmonium state, for example to
the $\psi (6S)$ \cite{Li:2009}, or interprete it as tetraquark state \cite{Cotugno:2010},
others see in it a $f_0(980)\psi '$ bound state \cite{Guo:2008}. For yet another
and may be somewhat unorthodox explanation see Ref.~\cite{vanBeveren:2009}.

In the present work we focus on the question whether the X(4660) and X(4630) could
be indeed one and the same state - and leave the issue of the dynamical origin aside.
While the background in the reaction
$e^+e^- \to \pi^+\pi^-\psi(2S)$ is fairly small and, therefore, one could argue that 
an extraction of the resonance parameters via a Breit-Wigner fit to the data
\cite{Lees:2014,Wang:2015} might be justified, this definitely cannot be said 
for $\eeLLc$. Due to the proximity of the $\LLc$ threshold (at $4573$ MeV) there is a
strong distortion of the signal and, clearly, the measured cross section
does not resemble a typical Breit-Wigner shape at all \cite{Belle08}.
Moreover, assuming that the transition is mediated by one-photon exchange the $\LLc$
will be either in the $^3S_1$ or $^3D_1$ partial wave. In the $S$ wave strong effects
from the final state interaction (FSI) are expected that will likewise influence
the energy dependence of the cross section. Such FSI effect arise from the coupling
to the resonance itself, but also from the residual interaction between the $\Lambda^+_c$ and
$\bar\Lambda^-_c$, say due to possible $t$-channel meson exchange, on top of an $s$-channel
resonance contribution.

The effects discussed above have been already considered in the arguments in
Refs.~\cite{Bugg:2009,Guo:2010} and are to some extent also simulated in the numerical
results presented there. However, since close to threshold a rather delicate interplay
between the resonance and the residual interaction (sometimes also called background
or non-pole contribution) has to be expected we believe that a more rigorous treatment
is required in order to obtain quantitatively reliable results and solid conclusions.
In recent studies of the reactions $\ee\to p\bar p$ \cite{Haidenbauer:2014} and
$\ee\to \Lambda\bar \Lambda$ \cite{Haidenbauer:2016} near their respective threshold we have
set up a framework that allows one to implement the FSI effects from the baryons in
a microscopic way. This formalism can be applied straight forwardly to the $\eeLLc$ case
as will be demonstrated in the present paper.
Though no resonances are present in the two reactions above, a clear enhancement
in the corresponding near-threshold cross sections has been found in
pertinent experiments. Our studies showed that a proper inclusion of the FSI effects within
our formalism allows one to achieve an excellent description of the measured cross sections.
In its application to $\eeLLc$ essential features such as the interplay between the
pole and non-pole part of the potential but also unitarity constraints on the $\LLc$
amplitude are implemented.
Moreover, a reliable extraction of the pole parameters of the X(4630) resonance is
possible, that does not rely on a Breit-Wigner parameterization, and these values can then
be confronted with the resonance properties extracted from the $e^+e^- \to \pi^+\pi^-\psi(2S)$ data.

The paper is structured as follows:
The ingredients of the $\LLc$ potential that is employed for generating the
FSI are summarized in Section~\ref{sec:2}. The potential involves
contact terms analogous to those that arise in chiral effective field
theory (EFT) up to next-to-leading order (NLO) and a contribution from a
(bare) resonance.
In addition, the relativistic Lippmann-Schwinger equation is introduced that is
solved in order to obtain the $\LLc \to \LLc$ amplitude, and the equation
for the distorted wave Born approximation that is used for calculating the
amplitude for the $\eeLLc$ transition.
In Section~\ref{sec:4} we describe our fitting procedure. The free parameters
in the $\LLc$ potential mentioned above are fixed in a fit to the
cross-section data for $\eeLLc$ by the Belle Collaboration \cite{Belle08}.
An excellent
reproduction of the experimental information can be achieved and is
presented in Section~\ref{sec:4} too. Furthermore, we extract the pole
position of the X(4630) that results from our fits and provide an estimate
for the uncertainty.
Finally, we summarize our results briefly in Section~\ref{sec:5}.


\section{Formalism}
\label{sec:2}

The principal features of the formalism employed in the present study of the reaction
$\eeLLc$ are identical to the one developed and described in detail in
Ref.~\cite{Haidenbauer:2014} where the reaction $\ee \to p \bar p$ was analyzed.
Therefore, we will be very brief here and focus primarily on aspects where there are
differences.

\subsection{The $\LLc$ interaction and the $\eeLLc$ transition amplitude}
\label{sec:3}

The $N \bar N$ interaction as needed for a calculation of the timelike electromagnetic
form factors of the proton near the $p \bar p$ threshold within the approach
outlined in Ref.~\cite{Haidenbauer:2014} is constrained by a wealth of empirical
information from $p\bar p \to p\bar p$ and $p\bar p \to n\bar n$ scattering experiments.
Specifically, there is a partial-wave analysis (PWA) available \cite{Zhou2012}.
Indeed, in our investigation \cite{Haidenbauer:2014} we utilized a $N\bar N$ potential derived
within chiral EFT \cite{Kang:2013}, fitted to the results of the PWA.
With regard to the timelike electromagnetic form factors of the $\Lambda$ the situation is
somewhat different. Here the only constraints for the $\lbarl$ force are provided by FSI
effects in the reaction $p\bar p \to \lbarl$. That reaction has been extensively investigated
in the PS185 experiment at LEAR and data are available for total and differential cross-sections
but also for spin-dependent observables \cite{PS185}. In our study of the reaction
$\ee\to\Lambda\bar \Lambda$ \cite{Haidenbauer:2016} we employed phenomenological $\Lambda\bar \Lambda$
potentials (based on meson-exchange) that were fitted to those PS185
data \cite{Haidenbauer:1991,Haidenbauer:1992}.

For the $\LLc$ interaction there are no empirical constraints from hadronic reactions. In principle,
one could follow the same strategy as done in Ref.~\cite{Haidenbauer:2010}
in an attempt to estimate the cross section for the reaction $p\bar p \to \LLc$ and invoke SU(4)
flavor symmetry to connect the $\LLc$ interaction with the one in the $\Lambda\bar \Lambda$ system,
see also Ref.~\cite{Wang:2016}. However, in the present study we want to avoid to make any such
basically phenomenological assumptions. Instead we aim at using the experimental information on the reaction
$\eeLLc$ itself to constrain and fix the interaction in the $\LLc$ system. We will see and
discuss below in how far this is possible.

In the actual construction of the $\LLc$ interaction we adopt chiral EFT \cite{Epelbaum:2008,EKM:2015}
as guide line and follow closely the procedure that has been already utilized in the derivation
of our $N\bar N$ interaction \cite{Kang:2013,DJM:2017}. In this framework the
potential is given in terms of pion exchanges and a series of contact interactions with an increasing
number of derivatives. The latter represent the short-range part of the baryon-baryon force and are
parameterized by low-energy constants (LECs), that need to be fixed in a fit to data.
Since we treat the reaction $\eeLLc$ in the one-photon exchange approximation, the $\LLc$ system 
can only be in the $^3S_1$ and $^3D_1$ partial waves. This limits rather strongly 
the number of LECs that need
to be determined. Note also that there is no contribution from one-pion exchange because $\Lambda^+_c$
($\bar \Lambda^-_c$) has isospin $I=0$. Given that the energy region of interest is in
the order of $100$ MeV we restrict ourselves to interactions up to NLO in the chiral expansion.
In principle, at NLO two-pion exchange contributions involving intermediate
$\Sigma_c\bar \Sigma_c$ states arise. However, in view of the rather large mass difference
$M_{\Sigma_c}-M_{\Lambda_c} \approx 167$ MeV we assume that such contributions can be effectively
absorbed into the contact terms.

The explicit form of the contact terms up to NLO is, after partial-wave projection~\cite{DJM:2017},
\begin{eqnarray}
V(^3S_1)(p',p)
&=& \tilde{C}_{^3S_1} + {C}_{^3S_1} ({p}'^2+{p}^2)\, -{\rm i}\, (\tilde C_{^3S_1}^a+C_{^3S_1}^ap'^2)\,(\tilde C_{^3S_1}^a+C_{^3S_1}^ap^2), \nonumber\\
V(^3D_1 -\, ^3S_1)(p',p) &=& {C}_{\epsilon_1}\, {p'}^2\, -{\rm i}\,C_{\epsilon_1}^a p'^2\, (\tilde C_{^3S_1}^a+C_{^3S_1}^ap^2), \nonumber \\
V(^3S_1 -\, ^3D_1)(p',p) &=& {C}_{\epsilon_1}\, {p}^2 \, -{\rm i}\,(\tilde C_{^3S_1}^a+C_{^3S_1}^ap'^2)\, C_{\epsilon_1}^a p^2, \nonumber \\
V(^3D_1)(p',p) &=& 0 \ ,
\label{VC}
\end{eqnarray}
with $p = |{\bf p}\,|$ and ${p}' = |{\bf p}\,'|$  the initial and final center-of-mass momenta
of the $\Lambda^+_c$ or $\bar \Lambda^-_c$.  Here, the $\tilde{C}_i$ denote the LECs that arise at LO and that
correspond to contact terms without derivatives, the ${C}_i$ arise at NLO from contact terms with two derivatives.
The term(s) right after the equality sign represent the elastic part of the $\LLc$ interaction.
The annihilation part is described likewise by contact terms but with a somewhat different form, in analogy
to the treatment of $N\bar N$ annihilation in our chiral EFT potential \cite{Kang:2013,DJM:2017}.
We refer the reader to Section 2.2 of Ref.~\cite{Kang:2013} for a thorough discussion and justification
for taking into account annihilation in this specific way. Here we just want to mention that the
choice is dictated primarily by the requirement to manifestly fulfil unitarity constraints on a formal level.
Note that in the expressions above the parameters $\tilde C^a$ and $C^a$ are real quantities. 

Since the Belle data suggest the presence of a resonance, the $X(4630)$ \cite{Belle08},
we include also a resonance in the $\LLc$ potential. It is done in form of a pole diagram
representing a bare vector-meson resonance with the quantum numbers $J^{PC}=1^{--}$ and $I=0$,
corresponding to a $\psi$-type $c\bar c$ meson.
Let us emphasize, however, that the introduction of such a pole diagram does not imply a
bias for the dynamical origin of this resonance which is still controversally
discussed in the literature \cite{Bugg:2009,Cotugno:2010,Guo:2010,vanBeveren:2009}. We are here only concerned
with the interplay of such a resonance structure (whatever its origin is) with the non-resonant part
of the $\LLc$ interaction and its consequences for the shape and the actual position of the (physical)
pole.

The potential is derived from the following Lagrangian that describes the coupling of a vector
meson to the $\Lambda_c$ ($\bar\Lambda_c$)  
\beq
{\cal L} = g_V \bar \Psi \gamma^\mu \Psi \phi_\mu
+ \frac{f_V}{4\,M_{\Lambda_c}} \bar \Psi \sigma^{\mu\nu} \Psi\, (\partial_\mu\phi_\nu-\partial_\nu\phi_\mu)
\ \ {\rm + \ \ H.c.} ~,
\label{Lag}
\eeq
with $\Psi$ and $\phi$ representing the fields of the $\Lambda^+_c$ and the vector meson, respectively.
The resulting potential after partial wave projection is of the form \cite{Hippchen:1989}
\bea
\label{VPOLE}
V_{^3S_1} (p',p;E) &=&\frac{4}{9 m_V(E-m_V)} 
\left[g_V\left(1+\frac{M_{\Lambda_c}}{2E_{p'}}\right)+f_V\left(\frac{E}{4M_{\Lambda_c}}+\frac{E}{2E_{p'}}\right)\right]
\,
\left[g_V\left(1+\frac{M_{\Lambda_c}}{2E_p}\right)+f_V\left(\frac{E}{4M_{\Lambda_c}}+\frac{E}{2E_p}\right)\right]\;,\no\\
V_{^3D_1} (p',p;E) &=& \frac{2}{9 m_V(E-m_V)} 
\left[g_V\left(1-\frac{M_{\Lambda_c}}{E_{p'}}\right)+f_V\left(\frac{E}{2E_{p'}}-\frac{E}{2M_{\Lambda_c}}\right)\right]
\,
\left[g_V\left(1-\frac{M_{\Lambda_c}}{E_p}\right)+f_V\left(\frac{E}{2E_p}-\frac{E}{2M_{\Lambda_c}}\right)\right]\;,\no\\
V_{^3D_1-^3S_1} (p',p;E) &=&\frac{2\sqrt{2}}{9 m_V(E-m_V)}
\left[g_V\left(1-\frac{M_{\Lambda_c}}{E_{p'}}\right)+f_V\left(\frac{E}{2E_{p'}}-\frac{E}{2M_{\Lambda_c}}\right)\right]
\,
\left[g_V\left(1+\frac{M_{\Lambda_c}}{2E_p}\right)+f_V\left(\frac{E}{4M_{\Lambda_c}}+\frac{E}{2E_p}\right)\right]\;,\no\\
V_{^3S_1-^3D_1} (p',p;E) &=&\frac{2\sqrt{2}}{9 m_V(E-m_V)} 
\left[g_V\left(1+\frac{M_{\Lambda_c}}{2E_{p'}}\right)+f_V\left(\frac{E}{4M_{\Lambda_c}}+\frac{E}{2E_{p'}}\right)\right]
\,
\left[g_V\left(1-\frac{M_{\Lambda_c}}{E_p}\right)+f_V\left(\frac{E}{2E_p}-\frac{E}{2M_{\Lambda_c}}\right)\right]\; ,\no\\
&&
\eea
where $E_p = \sqrt{p^2+M^2_{\Lambda_c}}$, $E_{p'}= \sqrt{p'^2+M^2_{\Lambda_c}}$, and $E=\sqrt{s}$
is the total energy. The quantity $m_V$ denotes the mass of the resonance, and $g_V$ and $f_V$ 
are the vector and tensor coupling constant, respectively. These are bare quantities and 
aquire their physical values by solving
the Lippmann-Schwinger equation, see below.

The coupling between the $\ee$ and $\LLc$ systems is constructed in close analogy to our treatment of
the photon coupling in pion photoproduction \cite{Ronchen:2014}. First we have a contact interaction,
which actually corresponds to the situation considered in our studies of $\ee \to p \bar p$ and
$\ee \to \Lambda\bar\Lambda$, and stands for a coupling via photon exchange.
In addition, a direct coupling of the $\ee$ pair to $\LLc$ via the bare resonance is included.
Thus, the Born amplitude for the transition $\eeLLc$ is described by
\bea
\label{EPOLE}
F^{0}_{^3S_1} (p',p;E) &=&-\frac{4\alpha}{9}\Biggl\{ G_{ee} \left(1+\frac{M_{\Lambda_c}}{2E_{p'}}\right) + 
\frac{g_{ee}}{m_V(E-m_V)}\,
\,\left[g_V\left(1+\frac{M_{\Lambda_c}}{2E_{p'}}\right)+f_V\left(\frac{E}{4M_{\Lambda_c}}+\frac{E}{2E_{p'}}\right)\right]\Biggr\}
\left(1+\frac{m_e}{2E_{p}}\right)\;,\no\\
F^{0}_{^3D_1} (p',p;E) &=&-\frac{2\alpha}{9}\Biggl\{ G_{ee} \left(1-\frac{M_{\Lambda_c}}{E_{p'}}\right) + 
\frac{g_{ee}}{m_V(E-m_V)}\,
\left[g_V\left(1-\frac{M_{\Lambda_c}}{E_{p'}}\right)+f_V\left(\frac{E}{2E_{p'}}-\frac{E}{2M_{\Lambda_c}}\right)\right]\Biggr\}
\left(1-\frac{m_e}{E_{p}}\right)\;,\no\\
F^{0}_{^3D_1-^3S_1} (p',p;E) &=&-\frac{2\sqrt{2}\alpha}{9}\Biggl\{
G_{ee}\left(1-\frac{M_{\Lambda_c}}{E_{p'}}\right) + \frac{g_{ee}}{m_V(E-m_V)}\,
\left[g_V\left(1-\frac{M_{\Lambda_c}}{E_{p'}}\right)+f_V\left(\frac{E}{2E_{p'}}-\frac{E}{2M_{\Lambda_c}}\right)\right]\Biggr\}
\left(1+\frac{m_e}{2E_{p}}\right)\;,\no\\
F^{0}_{^3S_1-^3D_1} (p',p;E) &=&-\frac{2\sqrt{2}\alpha}{9}\Biggl\{
G_{ee} \left(1+\frac{M_{\Lambda_c}}{2E_{p'}}\right)+ \frac{g_{ee}}{m_V(E-m_V)}\,
\left[g_V\left(1+\frac{M_{\Lambda_c}}{2E_{p'}}\right)+f_V\left(\frac{E}{4M_{\Lambda_c}}+\frac{E}{2E_{p'}}\right)\right]\Biggr\}
\left(1-\frac{m_e}{E_{p}}\right)\;.\no\\
&&
\eea
The quantities $G_{ee}$ and $g_{ee}$ represent the strengths of the coupling via a contact term and the bare 
resonance, respectively. The notation is chosen in such a way that the non-pole contribution in 
Eq.~(\ref{EPOLE}) matches the one in the corresponding Eq.~(6) of Ref.~\cite{Haidenbauer:2014}.

\subsection{Scattering equation}\label{sec:LSE}

The $\LLc$ amplitude is obtained from the solution of a relativistic Lippmann-Schwinger (LS) equation:
\begin{eqnarray}
T_{L''L'}(p'',p';E)&=&V_{L''L'}(p'',p';E)\nonumber\\
&+&
\sum_{L}\int_0^\infty \frac{dpp^2}{(2\pi)^3} \, V_{L''L}(p'',p;E)
\frac{1}{E-2E_p+i0^+}T_{LL'}(p,p';E),
\label{LS}
\end{eqnarray}
with $E=\sqrt{s}$. The potential $V$ is the sum of contact terms,
Eq.~(\ref{VC}), and the pole diagram, Eq.~(\ref{VPOLE}).
The scattering (on-shell) amplitude is given by
$T_{L''L'}(k):=T_{L''L'}(k,k;E)$, with $k$  the on-shell momentum defined by
$E = 2 E_k=2\sqrt{M^2_{\Lambda_c}+k^2}$.
In our study of the reaction $\eeLLc$ we restrict ourselves to the one-photon approximation
\cite{Haidenbauer:2014} so that we need only the coupled partial waves
$^3S_1$ and $^3D_1$, therefore $L'',L',L=0,2$.

The amplitude for the reaction $\eeLLc$ is evaluated in distorted wave Born approximation,
\begin{eqnarray}
F^{\LLc,\ee}_{L''L'}(k,k_e;E)&=&F^{0}_{L''L'}(k,k_e;E)\nonumber\\
&+&
\sum_{L}\int_0^\infty \frac{dpp^2}{(2\pi)^3} \, T_{L''L}(k,p;E)
\frac{1}{E-2E_p+i0^+} F^{0}_{LL'}(p,k_e;E)~,
\label{DWBA}
\end{eqnarray}
with $k_e$  the on-shell momentum of the $\ee$ pair and $E=2 E_k$. Here, $F^{0}_{L''L'}$ stands 
for the Born term for $\eeLLc$ as given in Eq.~(\ref{EPOLE}). 
Like the $\LLc$ potential itself, it depends explicitly on the energy $E$ because of the pole 
diagram, cf. Eq.~({\ref{EPOLE}).
From the amplitude $F^{\LLc,\ee}_{L''L'}$ the $\eeLLc$ cross section can be calculated in a
straightforward way, but also any other observable of the reaction $\eeLLc$,
see Ref.~\cite{Haidenbauer:2014}.

The potential $V$ that is inserted into the LS equation (\ref{LS}) needs to be regularized
in order to suppress high-momentum components \cite{Epelbaum:2008}. Following
Refs.~\cite{EKM:2015,DJM:2017} we do this by introducing a regulator function
with a cutoff mass. Since the contact interactions are non-local, cf. Eq.~({\ref{VC}), 
a non-local regulator is applied. Its explicit form is \cite{DJM:2017}
\begin{equation}
f(p',p) = {\rm exp}\left(-\frac{p'^m+p^m}{\Lambda^m}\right) \ .
\label{cutoff}
\end{equation}
In case of the transition potential for $\eeLLc$ only the momentum in the
$\LLc$ system acquires large values when evaluating Eq.~(\ref{DWBA}) and,
therefore, the corresponding contributions are likewise cut off.
For the cutoff mass $\Lambda$ we consider a range similar to the one 
regarded in Ref.~\cite{DJM:2017}. Specifically, we employ values between $0.45$~GeV 
and $0.85$~GeV.
Following \cite{EKM:2015}, the exponent in the regulator is chosen to be $m=2$.

We use the $\Lambda^+_c$ mass $M_{\Lambda_c} = 2286.46$ MeV \cite{PDG2016} so that
the $\LLc$ threshold is at $\sqrt{s} = 4572.92$ MeV.
As in Ref.~\cite{Haidenbauer:2014} we neglect the Coulomb interaction between the
$\Lambda^+_c$ $\bar\Lambda^-_c$ when solving the LS equation but include its
effect via the Sommerfeld-Gamow factor in the evaluation of the cross section.
In general, we use the speed plot to determine the pole position. However,
for the case of an elastic $\LLc$ interaction one can determine the pole also
by an analytical continuation of the $T$ matrix to the second Riemann sheet,
by exploiting that zeros of the $S$-matrix on the first sheet correspond to
poles on the second sheet. Doing so we can check the reliability of the results
obtained from the speed plot.

\section{Results}
\label{sec:4}

\subsection{Fitting procedure}

The parameters of the $\LLc$ potential are determined in a fit to the $\eeLLc$ cross section
of the Belle Collaboration \cite{Belle08}. This concerns the LECs, see Eq.~(\ref{VC}),
but also the bare parameters of the resonance, $m_V$, $g_V$, and $f_V$.
In the fit we consider data up to a kinetic center-of-mass energy of $100$ MeV in the $\LLc$ system,
which corresponds to $\sqrt{s}\leq 4.68$~GeV. Based on our experience with $\ee\to p\bar p$ and
$\ee \to \Lambda\bar\Lambda$, we expect the (electro-magnetic) couplings to the $\ee$ system 
($G_{ee}$, $g_{ee}$) to be practically constant over that energy range so that they amount just 
to normalization factors. 
With the above choice the data set comprises the first 6 points from Belle. However,
since the point at the lowest energy is below the nominal $\LLc$ threshold it is not explicitly
included in the least square minimization. Here we only make sure that our result at the 
threshold lies well within the pertinent bin.
Note that the cross section for $\eeLLc$ remains finite even at the $\LLc$ threshold because of
the attractive Coulomb interaction between $\Lambda^+_c$ and $\bar\Lambda^-_c$, see the analogous
situation for the $p\bar p$ final state \cite{Haidenbauer:2014}.

For the analysis of the Belle data we consider a variety of fit scenarios. First of all, we explore in how far
our results depend on the regularization procedure. For that we perform fits for a selection of cutoff masses
between $0.45$ and $0.85$ GeV, so that we cover an even wider range as considered in the 
$NN$ \cite{EKM:2015} and $N\bar N$ \cite{DJM:2017} studies.
We perform also fits with a different number of contact terms in the $\LLc$ interaction,
starting from a LO elastic $\LLc$ potential (one contact term, $\tilde C_{^3S_1}$) up to NLO and including
an elastic part as well as annihilation (four contact terms,
$\tilde C_{^3S_1}$, $\tilde C^a_{^3S_1}$, $C_{^3S_1}$, $C^a_{^3S_1}$).
Finally, we consider the cases where the $\ee$ state couples to the $\LLc$ system only via the resonance
and where it couples also directly via the photon, which corresponds to a contact interaction in our
formalism.

In exploratory fits we included also the contact terms $C_{\varepsilon_1}$, $C^a_{\varepsilon_1}$ that
introduce a $^3S_1$-$^3D_1$ coupling. However, it turned
out that the Belle data \cite{Belle08} do not allow one to fix those terms and results with or
without them were practically indistinguishable. Thus, we set them to zero. The same is also the 
case with the tensor coupling constant $f_V$ of the pole diagram, cf. Eq.~(\ref{VPOLE}),
so that we put $f_V=0$ in our analysis.

\begin{table}[h!]
\begin{center}
{\footnotesize
 \vspace{0.3cm}
\renewcommand{\arraystretch}{1.2}
\begin{tabular}{|c c c c c c c|}
\hline\hline
$\Lambda$ (GeV)                 &0.45  & 0.50    &   0.55     &  0.65     &  0.75     &  0.85    \\
\hline
\hline
 \multicolumn{7}{|l|}{with pole term, see Eq.~(\ref{EPOLE})}     \\
\hline
$\tilde{C}_{^3S_1}$ (GeV$^{-2}$)   & 191.8   & 110.1  & 61.27  & 7.853  & $-$19.17  &$-$34.48   \\
$g_V$                           & $-$8.734  &$-$8.123  &$-$7.625  &$-$6.837  &$-$6.218   &$-$5.706   \\
$m_V$~(GeV)                     & 4.6344  & 4.6364 & 4.6383 & 4.6419 & 4.6448   & 4.6472  \\
$g_{ee}$($\times10^{-3}$GeV$^2$) & 1.052  & 1.067  & 1.081  & 1.102  & 1.116    & 1.126   \\
$\chi^2$                        & 0.1          & 0.3        &  0.4    & 0.6     & 0.7     & 0.8       \\
pole (GeV)                      & 4.6550      & 4.6534     & 4.6514   & 4.6482    & 4.6462    & 4.6451     \\
                                & $-${\rm i}\,0.0264 & $-${\rm i}\,0.0311 & $-${\rm i}\,0.0343 & $-${\rm i}\,0.0376 & $-${\rm i}\,0.0389 & $-${\rm i}\,0.0394 \\
$a$ (fm)                        & $-$0.269      &$-$0.485      &$-$0.634      & $-$0.818    &$-$0.927     & -1.002        \\
\hline
\hline
 \multicolumn{7}{|l|}{with pole and non-pole contribution, see Eq.~(\ref{EPOLE})}     \\
\hline
$\tilde{C}_{^3S_1}$ (GeV$^{-2}$)   & 191.9   & 111.9  & 65.65 &$-$0.0100    &$-$11.76    &$-$26.96    \\
$g_V$                           & $-$8.808 &$-$7.964  &$-$7.356 &$-$6.490      &$-$5.899    &$-$5.415    \\
$m_V$~(GeV)                     & 4.6328  & 4.6398  & 4.6443  & 4.6473   & 4.6542 & 4.6572    \\
$g_{ee}$($\times10^{-3}$GeV$^2$)& 1.055  & 1.052   & 1.042  & 1.045    & 1.004    & 0.987   \\
$G_{ee}$($\times10^{-3}$)    & 0.272     &$-$0.578  &$-$1.035 &$-$1.100   & $-$1.672     & $-$1.787   \\
$\chi^2$                        & 0.1         & 0.1         &  0.2       & 0.2          & 0.2        & 0.2        \\
pole (GeV)                      & 4.6543      & 4.6552      & 4.6554     & 4.6532       & 4.6550     & 4.6549     \\
& $-${\rm i}\,0.0276 & $-${\rm i}\,0.0284 & $-${\rm i}\,0.0295 & $-${\rm i}\,0.0304 & $-${\rm i}\,0.0314 & $-${\rm i}\,0.0319 \\
$a$ (fm)                        & $-$0.325      &$-$0.360      &$-$0.403      & $-$0.641    &$-$0.538     &$-$0.581        \\
\hline
\hline
\end{tabular}
\caption{\label{tab:chi1}
Parameters of the fit at LO and without annihilation, for different cutoff masses $\Lambda$.
The given $\chi^2$ is for the data points below $\sqrt{s} = 4.68$~GeV, see text.
The $\LLc$ scattering length in the $^3S_1$ partial wave is denoted by $a$.
}
}
\end{center}
\renewcommand{\arraystretch}{1.0}
\end{table}

\begin{figure}[htbp]
\centering
\includegraphics[width=0.48\textwidth,height=0.30\textheight]{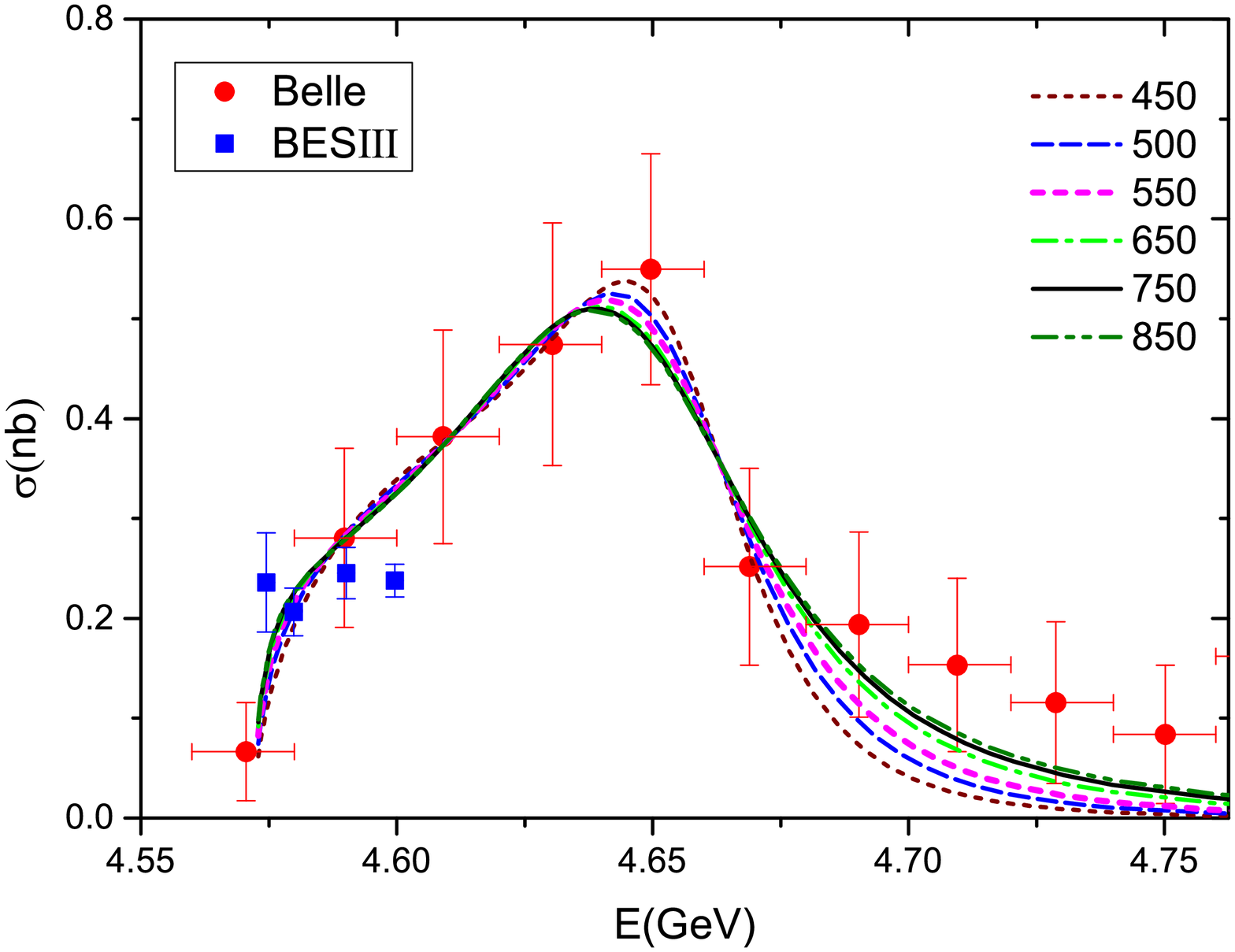}
\includegraphics[width=0.48\textwidth,height=0.30\textheight]{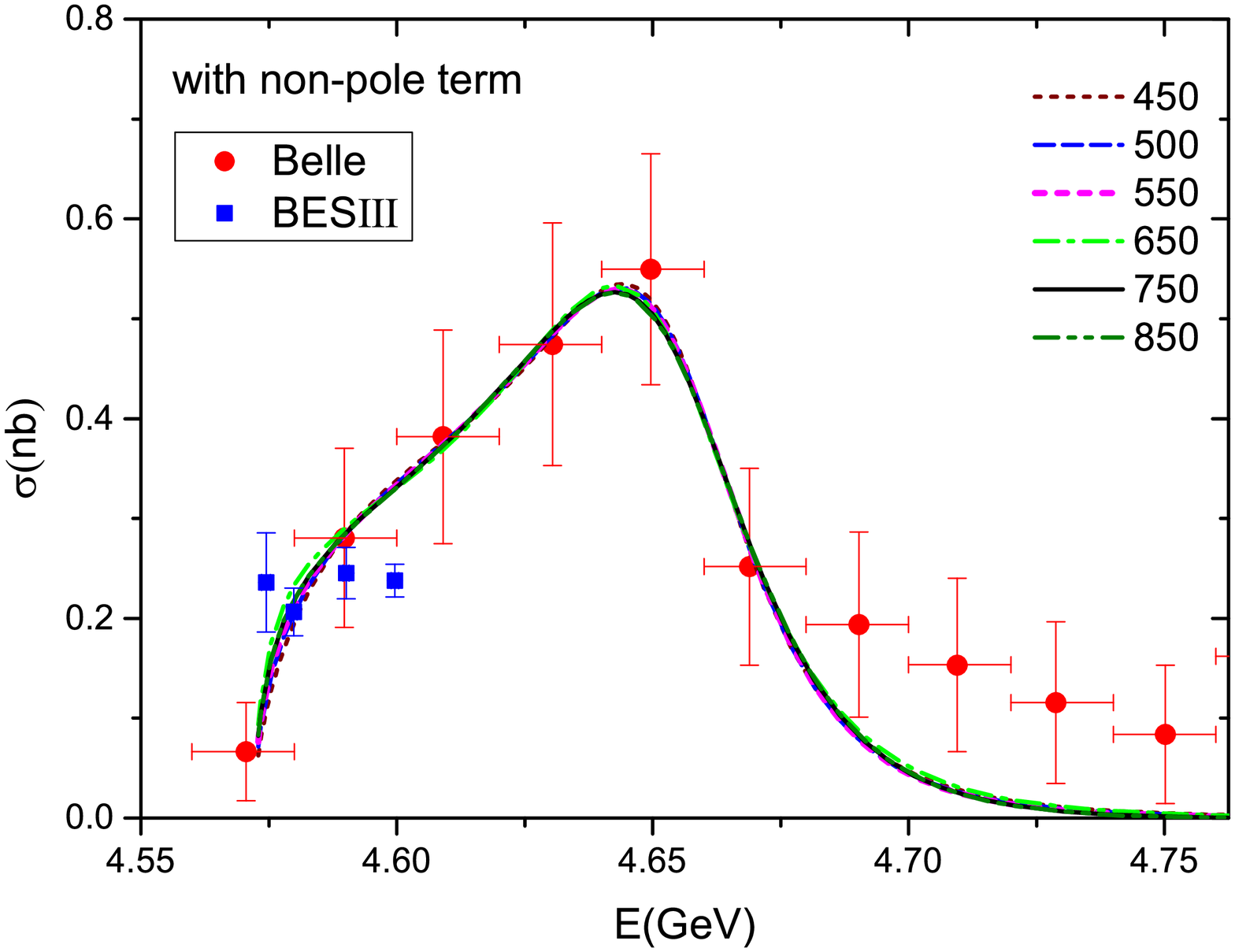}
\caption{Fits to the $\eeLLc$ cross section of Belle~\cite{Belle08}
(red circles)
for various cutoff masses $\Lambda$ at LO and without annihilation.
Left: Coupling between $\ee$ and $\LLc$ only via pole term.
Right: Coupling between $\ee$ and $\LLc$ via pole term plus non-pole term,
cf. Eq.~(\ref{EPOLE}).
The data from BESIII \cite{BESIII} (blue squares) are included for
illustration.
\label{fig:cs1}
}
\end{figure}

In a first series of fits we included only the contact term $\tilde C_{^3S_1}$, corresponding to
a purely elastic $\LLc$ potential at LO, together with the pole diagram and varied the cutoff
mass $\Lambda$. The resulting cross sections are displayed in Fig.~\ref{fig:cs1} for the
cases where the $\ee$ system couples either only via the resonance to $\LLc$ (left side)
or also via a contact term (right side).
The numerical values of the parameters are compiled in Table \ref{tab:chi1}.
In a second series of fits we added more and more terms in the contact
interaction, allowing not only for elastic scattering but also for
annihilation in the $\LLc$ channel. Here the cutoff mass is kept the same
for all interactions and fixed to $\Lambda= 0.75$~GeV.
The resulting cross sections are displayed in Fig.~\ref{fig:cs2}, again for
the cases where the $\ee$ system couples either only via the resonance
to $\LLc$ (left side) or also via a contact term (right side).
The numerical values of the parameters are compiled in Table \ref{tab:chi2}.

\begin{figure}
\centering
\includegraphics[width=0.48\textwidth,height=0.30\textheight]{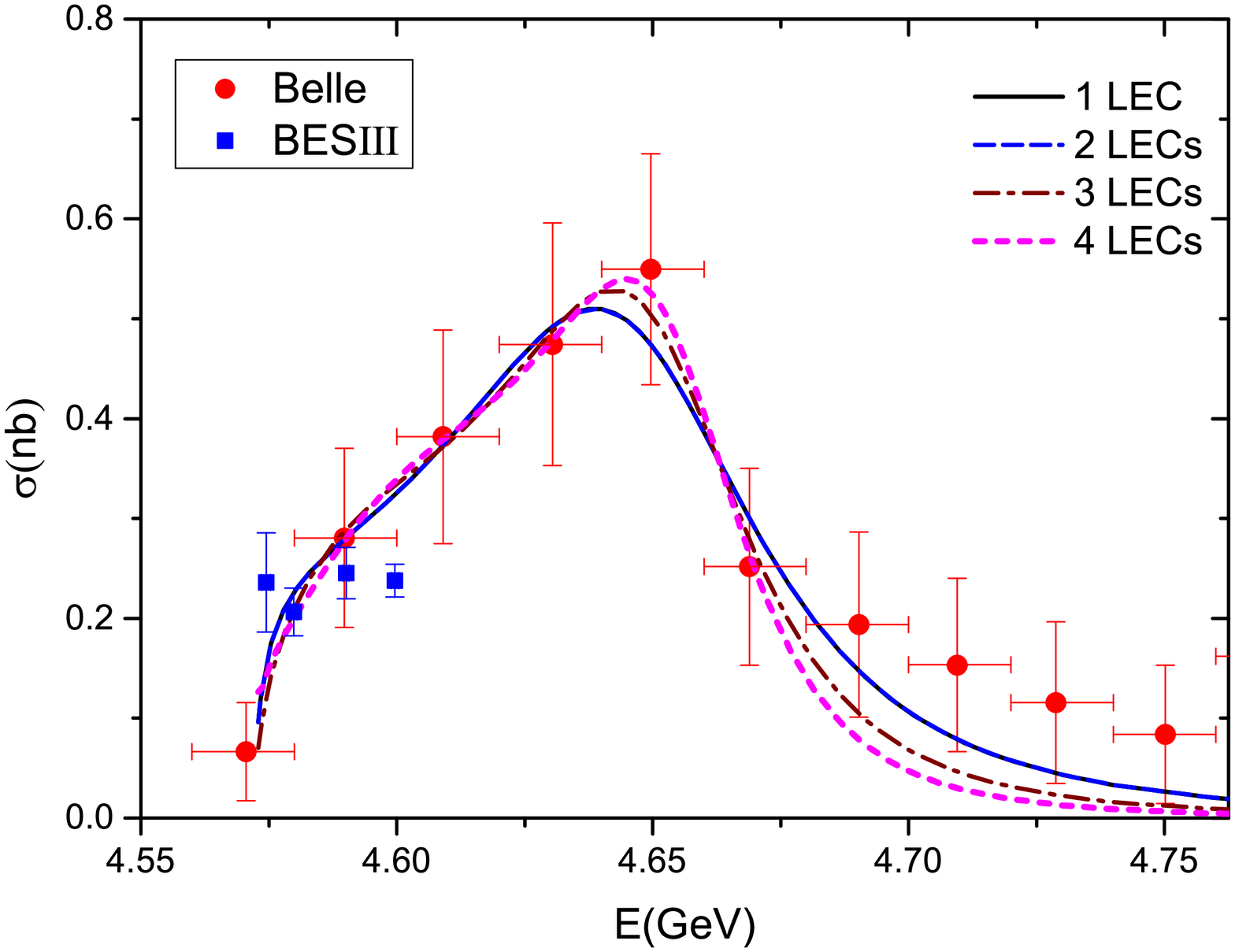}
\includegraphics[width=0.48\textwidth,height=0.30\textheight]{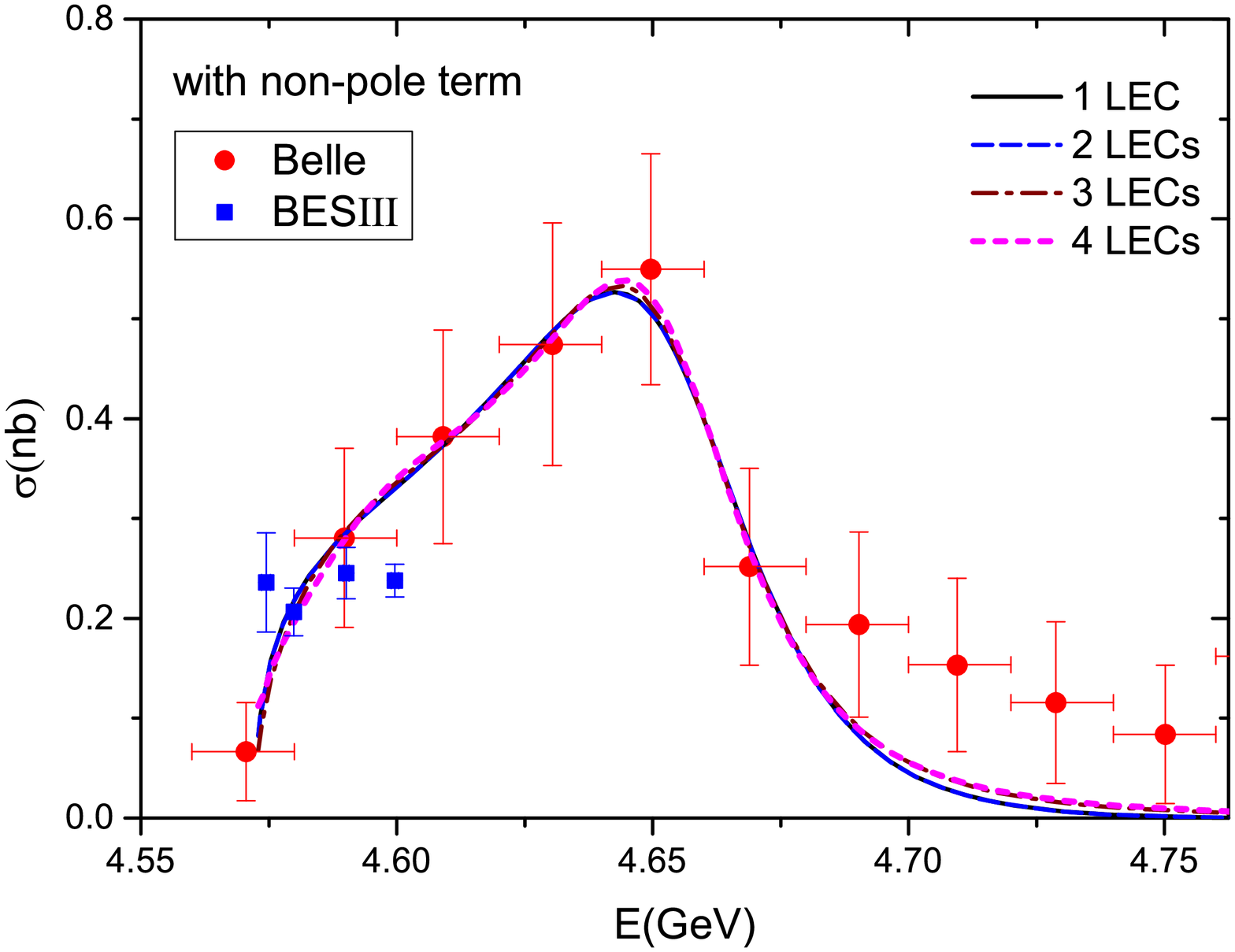}
\caption{Fits to the $\eeLLc$ cross section of Belle~\cite{Belle08}
(red circles)
at LO, without (1 LEC) and with annihilation term (2 LECs),
and up to NLO, without (3 LECs) and with annihilation term (4 LECs).
Left: Coupling between $\ee$ and $\LLc$ only via pole term.
Right: Coupling between $\ee$ and $\LLc$ via pole term plus non-pole term,
cf. Eq.~(\ref{EPOLE}).
The data from BESIII \cite{BESIII} (blue squares) are included for
illustration.
\label{fig:cs2}
}
\end{figure}

\begin{table}[h!]
\begin{center}
{\footnotesize
 \vspace{0.3cm}
\renewcommand{\arraystretch}{1.2}
\begin{tabular}{|c| c| c| c| c|  }
\hline\hline
                                   &  1 LEC          & 2 LECs      &  3 LECs     & 4 LECs        \\
\hline
 \multicolumn{5}{|l|}{with pole term, see Eq.~(\ref{EPOLE})}     \\
\hline
$\tilde{C}_{^3S_1}$~(GeV$^{-2}$)   &-19.17           &-19.23       &-0.1001        &-49.78         \\
$C_{^3S_1}$~(GeV$^{-4}$)           & -               & -           &-191.3         &-146.4      \\
$\tilde{C}^a_{^3S_1}$~(GeV$^{-1}$) & -               &0.1661       &-0.5353        &-1159        \\
$C^a_{^3S_1}$~(GeV$^{-3}$)         & -               & -           & -             & 4567        \\
$g_V$                              &-6.218           &-6.218       &-5.071         &-4.705          \\
$m_V$~(GeV)                        & 4.6448          & 4.6448      & 4.6386        & 4.6362          \\
$g_{ee}$($\times10^{-3}$GeV$^2$)   & 1.116           & 1.116       & 1.079         & 1.171         \\
$\chi^2$                           & 0.7           & 0.7       &  0.3        &  0.1           \\
pole (GeV) &$4.6462-{\rm i}\,0.0389$ & $4.6455-{\rm i}\,0.0390$ & $4.6501-{\rm i}\,0.0396$ & $4.6506-{\rm i}\,0.0397$    \\
$a$ (fm)                          &$-0.927$         & $-0.928$    & $-0.726$      & $-0.916-{\rm i}\,0.844$    \\
\hline
\hline
 \multicolumn{5}{|l|}{with pole and non-pole contribution, see Eq.~(\ref{EPOLE})}     \\
\hline
                                  &  1 LEC       & 2 LECs       &  3 LECs        & 4 LECs        \\
\hline
$\tilde{C}_{^3S_1}$~(GeV$^{-2}$)  &-11.76        &-11.74   )    &-0.0135         &-60.76         \\
$C_{^3S_1}$~(GeV$^{-4}$)          & -            & -            &-187.9          & -74.23         \\
$\tilde{C}^a_{^3S_1}$~(GeV$^{-1}$)& -            & 0.6595          & 0.0503         & -1185        \\
$C^a_{^3S_1}$~(GeV$^{-3}$)        & -            & -            & -              & 5455          \\
$g_V$                             &-5.899         &-5.897       &-5.012          &-4.858          \\
$m_V$~(GeV)                       & 4.6542           & 4.6542     & 4.6414       & 4.6342        \\
$g_{ee}$($\times10^{-3}$GeV$^2$)&1.004          & 1.003       & 1.063          & 1.200     \\
$G_{ee}$($\times10^{-3}$)      &-1.672         &-1.679       &-0.455          &0.512               \\
$\chi^2$                          & 0.2              & 0.2        &  0.2        &  0.1           \\
pole (GeV)& $4.6550-{\rm i}\,0.0314$ & $4.6546-{\rm i}\,0.0312$  & $4.6520-{\rm i}\,0.0285$  & $4.6482-{\rm i}\,0.0341$    \\
$a$ (fm)                          &$-0.538$            & $-0.537$   & $-0.632$          & $-0.981-{\rm i}\,0.714$    \\
\hline
\hline
\end{tabular}
\caption{\label{tab:chi2}
Parameters of the fits up to NLO, with/without annihilation term. The cutoff mass $\Lambda$ is $0.75$~GeV.
The given $\chi^2$ is for the data points below $\sqrt{s} = 4.68$ GeV, see text.
The $\LLc$ scattering length in the $^3S_1$ partial wave is denoted by $a$.
}
}
\end{center}
\renewcommand{\arraystretch}{1.0}
\end{table}

\subsection{Discussion of results}

The results presented in Figs.~\ref{fig:cs1} and \ref{fig:cs2} attest that the
Belle data can be reproduced rather well over the fitting range within all scenarios
considered. Differences in the cross sections appear mainly at higher energies.
There is also some variation around the maximum, where the fits that include
a non-pole term in the electromagnetic coupling reproduce the peak value and
the subsequent sharp drop in the cross section visibly better.
Note that in the course of our study
we have also performed extended fits where all data points up to $4.75$~GeV
were included (though by giving less weight to the data at higher energies).
Those led to results that are practically identical to the ones shown in
Figs.~\ref{fig:cs1} and \ref{fig:cs2}.

Let us discuss the results more thoroughly and, to begin with, look at the cutoff 
dependence. 
There are still noticeable variations in the scenario where only the coupling via 
a pole term is considered (upper part of Table~\ref{tab:chi1}). Specifically, 
there is an observable deterioration in the achieved $\chi^2$ with increasing
cutoff mass. Moreover, there is a pronounced variation of the $\LLc$ $^3S_1$
scattering length $a$. On the other hand, the resonance parameters
themselves are less sensitive to the cutoff. The variations of the
resonance parameters, given in terms of the real and imaginary part of the 
pole position in Table~\ref{tab:chi1}, are in the order of $10$ MeV or so. 
Evidently, once a non-pole contribution is added the cutoff dependence is 
remarkably reduced, cf. the lower part of Table~\ref{tab:chi1}. 
First, now the achieved $\chi^2$ is practically the same for all cutoffs.
The variation in $a$ is much smaller and, actually, within the expected 
uncertainty for the determination of the scattering length from an FSI analysis 
estimated in Ref.~\cite{Gasparyan:2003} on general grounds. Finally, the variation 
in the resonance mass is only about $2$ MeV, and around $8$ MeV for the width. 
We interprete these variations as the inherent systematic error of our analysis. 

Results considering variations of the $\LLc$ interaction are summarized in
Table~\ref{tab:chi2}. Since the influence of the cutoff has been established
above, we show only results for a fixed cutoff value, namely for 
$\Lambda = 0.75$~GeV. 
Again, fits that include either a pole term alone or a pole and a non-pole 
coupling to $\ee$ have been performed. However, in view of the preceding 
discussion we expect primarily the latter scenario to provide reliable and 
physically meaningful results. Indeed, again practically the same $\chi^2$ 
could be achieved, independendly of whether just a single term (elastic) $\LLc$
interaction is employed or one with 4 LECs that involves contributions to 
the elastic part and annihilation up to NLO. 
Actually, now also the resulting scattering lengths are fairly close together,
at least for the first three $\LLc$ potentials. Only for the one with 
4 LECs there is a striking difference. It has to be said, however, that
in this particular fit we have tried intentionally to increase annihilation 
as much as possible - in order to explore possible consequences for the 
resulting scattering length but also the pole position.  
As such, this exercise reveals that the $\eeLLc$ cross section data do not 
allow a unique determination of the $\LLc$ interaction. However, in view
of the presence of annihilation in the $\LLc$ channel this is not really a surprise. 
 
Fortunately, the resonance parameters are much less sensitive to details of
the $\LLc$ interaction and, specifically, to the strength of annihilation, 
cf. the corresponding results in the lower part of Table~\ref{tab:chi2}.
Utilizing these variations as basis for estimating the uncertainty of the 
resonance parameters of the X(4630) we arrive at
$M = (4652.5\pm 3.4)$~MeV and $\Gamma = (62.6\pm 5.6)$~MeV.
These values have to be compared with the ones from the Belle fit which are 
$M= 4634^{+8}_{-7}{}^{+5}_{-8}$ MeV and $\Gamma = 92^{+40}_{-24}{}^{+10}_{-21}$ MeV \cite{Belle08}.
Though our results agree with the ones of Belle within the given uncertainties, the central
value of the resonance mass extracted from our analysis is clearly shifted upwards by about
$20$ MeV as compared to the one from the Breit-Wigner fit, while the width is signficantly 
smaller.
The latest results for the X(4660) from measurements of the $\pi^+\pi^-\psi(2S)$ channel are
$M = (4652\pm 10\pm 8)$~MeV and $\Gamma = (68\pm 11\pm 1)$~MeV (Belle \cite{Wang:2015}), and
$M = (4669\pm 21\pm 3)$~MeV and $\Gamma = (104\pm 48\pm 10)$~MeV
(BaBar \cite{Lees:2014}).
Obviously, there is a remarkable agreement between our X(4630) parameters determined
from $\eeLLc$ data with the ones extracted by Belle for the X(4660) in the
$\ee \to \pi^+\pi^-\psi(2S)$ decay. The X(4660) parameters given by BaBar are somewhat
different, but one has to take into consideration that the uncertainties are much larger in
the latter determination.
An overview of the resonance parameters is provided in Table~\ref{tab:width}. 

\begin{table}[h!]
\begin{center}
{\footnotesize
 \vspace{0.3cm}
\renewcommand{\arraystretch}{1.6}
\begin{tabular}{|c|c|c|c|c|  }
\hline\hline
             &    present analysis      &    Belle \cite{Belle08}  &  Belle \cite{Wang:2015} & BABAR \cite{Lees:2014} \\
\hline
reaction     &    $\eeLLc$ & $\eeLLc$  &  $\ee\to\pi^+\pi^-\psi(2S)$ & $\ee\to\pi^+\pi^-\psi(2S)$ \\
\hline
mass  $M$ (MeV)      & $4652.5\pm 3.4\pm 1.1$ & $4634^{+8}_{-7}{}^{+5}_{-8}$   & $4652\pm 10\pm 8$ & $4669\pm 21\pm 3$ \\
width $\Gamma$ (MeV) & $62.6\pm 5.6\pm 4.3$   & $92^{+40}_{-24}{}^{+10}_{-21}$ & $68\pm 11\pm 1$ & $104\pm 48\pm 10$  \\
\hline
\end{tabular}
\caption{\label{tab:width}
Overview of resonance parameters for the X(4630) and X(4660), respectively. 
}
}
\end{center}
\renewcommand{\arraystretch}{1.0}
\end{table}

We do not include the $\LLc$ invariant mass spectrum measured in the
reaction $\bar B \to \LLc \bar K$ \cite{Aubert:2008} in our fit.
Given that $M_B = 5279$ MeV and $2M_{\Lambda_c}+M_K \approx 4948$ MeV
the phase space for the decay $\bar B \to \LLc \bar K$ is fairly small.
Because of that it is likely that the $\LLc$ spectrum is significantly
distorted by possible interactions in the other subsystems, $\Lambda^+_c K^-$ and/or
$\bar\Lambda^-_c K^-$.  Indeed, the invariant mass spectrum for $\Lambda^+_c K^-$
shown in Ref.~\cite{Aubert:2008} suggests the presence of a $\Xi_c$ resonance in that 
channel around $2930$ MeV. See also the related discussion in Ref.~\cite{Guo:2008}.
Further complications for an analyis are the relatively low statistics of the
data and the fact that $\LLc$ FSI effects could come not only from the $^3S_1$ but
also from the $^1S_0$ partial wave, because parity is not conserved
in this decay so that the $\bar K$ can be in an $s$- or $p$ wave.

\begin{figure}
\centering
\includegraphics[width=0.48\textwidth,height=0.30\textheight]{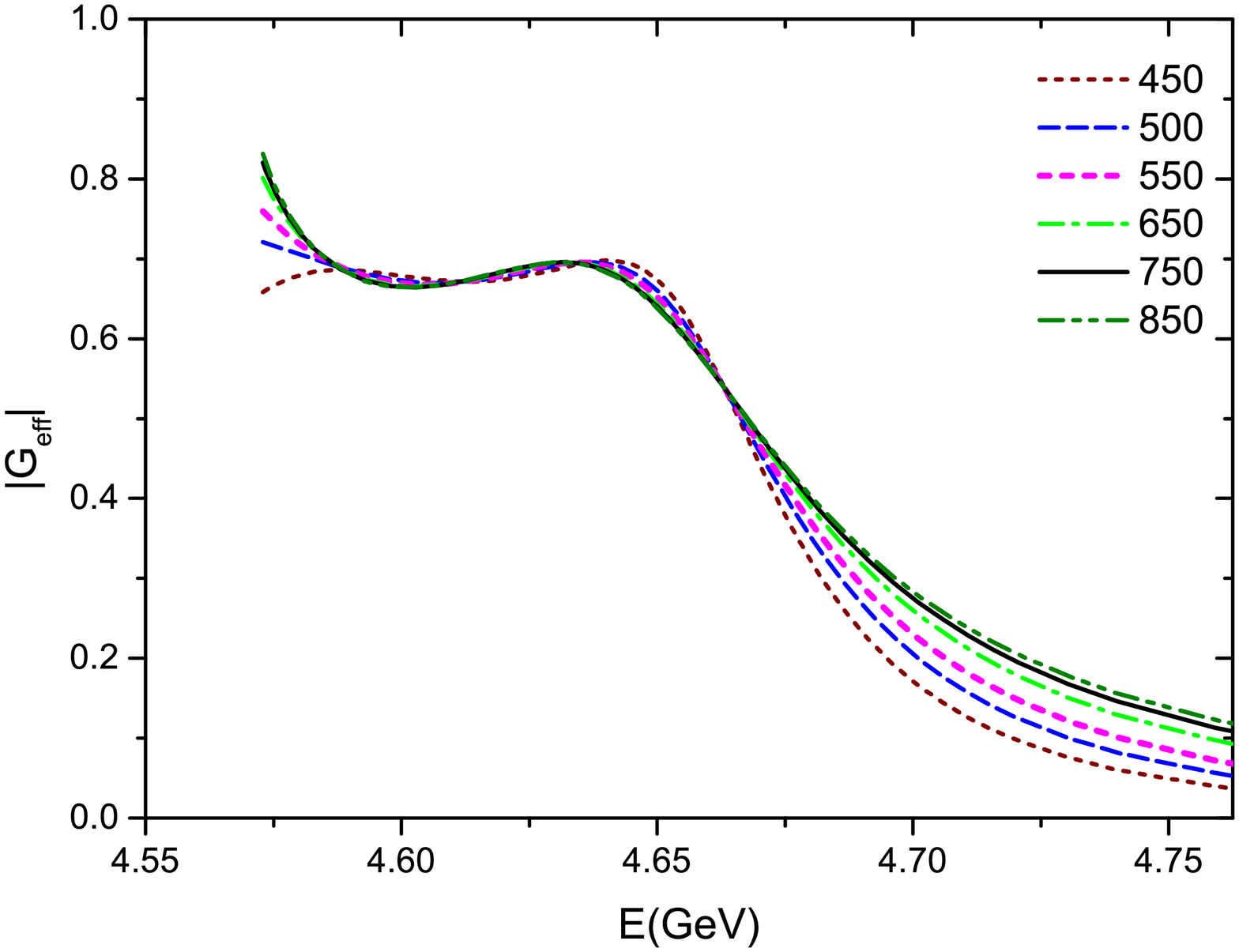}
\includegraphics[width=0.48\textwidth,height=0.30\textheight]{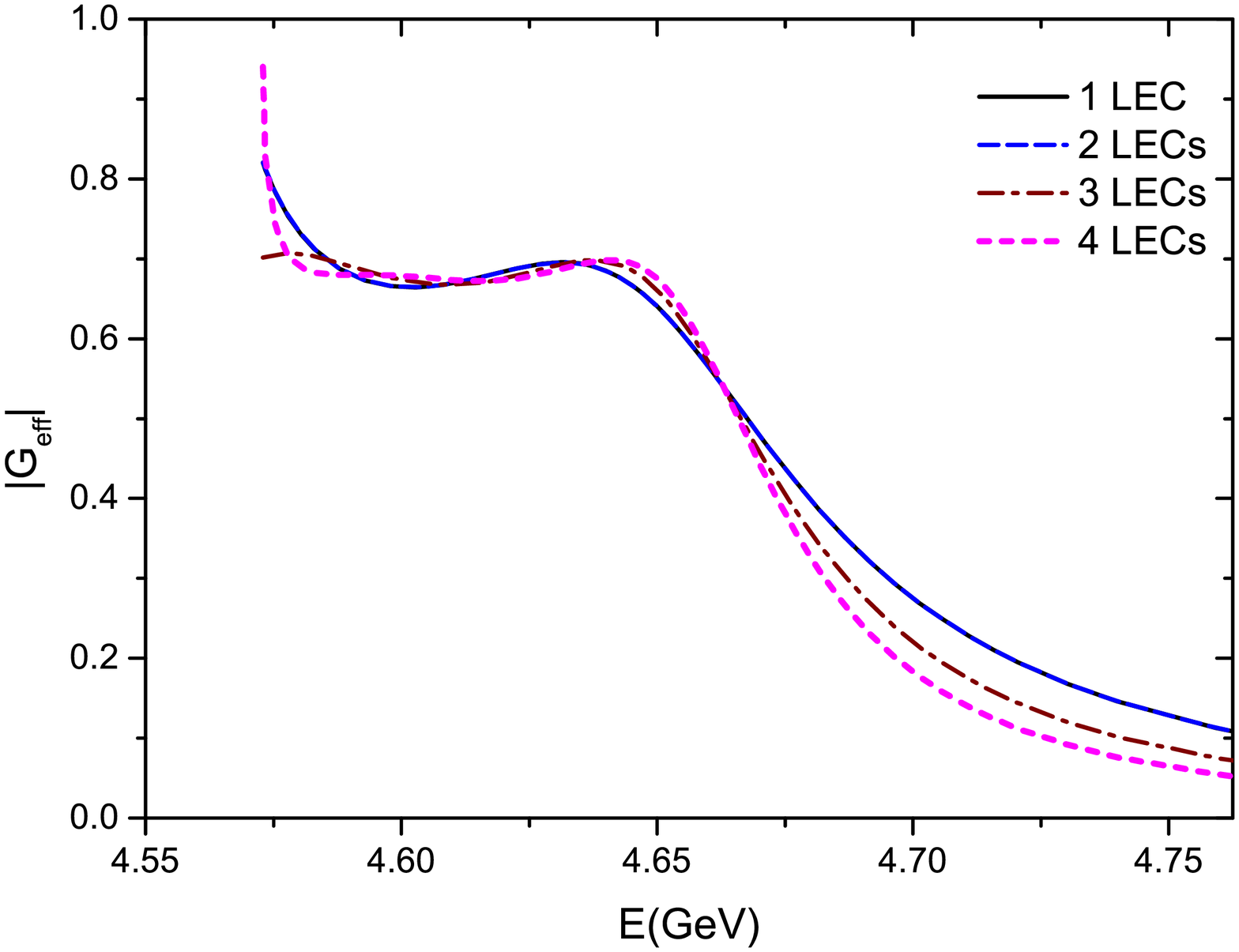}
\caption{Prediction for the effective form factor $G_{\rm eff}$.
Left panel: LO results for various values of the cutoff $\Lambda$.
Right panel:  Results at LO and NLO, with/without annihilation term. The cutoff mass $\Lambda$ is $0.75$~GeV.
For a detailed description of the employed $\LLc$ interactions, see text. 
}
\label{fig:geff}
\end{figure}

\subsection{Outlook on the $\Lambda_c$ electromagnetic form factors}

One of the motivations for measurements of reactions like
$\ee\to p\bar p$ and $\ee\to \Lambda\bar \Lambda$ is that one can determine
the electromagnetic form factors of the corresponding baryons in the time-like
region~\cite{Denig}. This applies also to the $\Lambda^+_c$. Indeed, recently a
new measurement of the reaction $\eeLLc$ has been performed by the BESIII
Collaboration~\cite{BESIII} and first results for the ratio of the $\Lambda^+_c$
electromagnetic form factors $G_E$ and $G_M$ have been presented.

We include the cross section data from the BESIII measurement in Figs.~\ref{fig:cs1}
and \ref{fig:cs2} for illustration. However, we want to emphasize that they
were not taken into account in our analysis of the $X(4630)$, which is the main
goal of the present paper.
While these data agree with the ones from the Belle Collaboration \cite{Belle08}
as far as the magnitude of the reaction cross section is concerned, they seem
to indicate a different trend for the energy dependence. Exploratory fits
with inclusion of those data revealed that it is practically imposible to
reconcile this trend with the Belle data at energies around the $X(4630)$
peak based on a $\LLc$ FSI that is constructed along the lines of chiral EFT, 
see Eqs.~(\ref{VC}) and (\ref{VPOLE}).
Hopefully, the BESIII Collaboration will be able to extend their measurements
to somewhat higher energies and, thereby, clarify the situation.
If the trend suggested by the BESIII data (cf. Figs.~\ref{fig:cs1} and
\ref{fig:cs2}) persists even for energies closer to the $X(4630)$, it will have
a drastic impact on the actual parameters of the resonance.
Anyway, in anticipation of future results from BESIII, predictions for the
effective electromagnetic form factor of the $\Lambda_c$ are displayed in
Fig.~\ref{fig:geff}, for the fits where the $\ee$ pair couples to $\LLc$ via the
pole term alone. Results for the variants where a non-pole coupling is included
are very similar and, therefore, not shown.
For the definition of $G_{\rm eff}$ see, e.g., Ref.~\cite{Haidenbauer:2014}.

There are also results for the angular distribution of the $\Lambda_c$ in Ref.~\cite{BESIII}.
The data are for $\sqrt{s}=4.5745$ GeV and $\sqrt{s}=4.5995$ GeV,
respectively, corresponding to kinetic energies of $1.6$ MeV and $26.6$ MeV in
the $\LLc$ system. At the lower energy the angular distribution is rather flat suggesting
that the $\LLc$ state is produced almost entirely in the $^3S_1$ partial wave.
This behavior is well in line with our calculation. At the higher
energy the data indicate the presence of contributions from the $^3D_1$
partial wave. Thus, in a future analysis one could use those data to fix the
additional LECs ($C_{\varepsilon_1}$, $C^a_{\varepsilon_1}$) in our NLO interaction,
see Eq.~(\ref{VC}), which could not be determined from the Belle data, as discussed in Section 3.1.
Also here results at higher energies would be rather helpful in order to map out the
actual energy dependence of the $D$-wave contribution.

\section{Summary}
\label{sec:5}

In the present work we investigated the reaction $\eeLLc$ at energies close to
the threshold with the aim to examine the impact of the $X(4630)$ resonance and to
determine its parameters. Thereby, special emphasis was put on a rigorous
treatment of the interaction in the final $\LLc$ state. The latter was done in
distorted wave Born approximation, following our works on
$\ee \to p \bar p$ \cite{Haidenbauer:2014}
and $\ee \to \Lambda \bar \Lambda$ \cite{Haidenbauer:2016}.

The relevant interaction in the $\LLc$ system was constructed along the lines of
chiral effective field theory up to next-to-leading order, supplemented by a 
pole diagram that represents a bare $X(4630)$ resonance.
The inherent parameters (low-energy constants, bare mass and coupling constant
of the resonance) were determined in a fit to the $\eeLLc$  data of the
Belle Collaboration \cite{Belle08}. Since it turned out that a unique determination
of involved parameters in a fit to these data is not possible we considered a variety of
scenarios in order to estimate the uncertainty of the results for the $X(4630)$ resonance.
Based on those variants the pole parameters of the $X(4630)$ were found to be
$M = (4652.5\pm 3.4\pm 1.1)$~MeV and $\Gamma = (62.6\pm 5.6 \pm 4.3)$~MeV,
where the first uncertainty is due to variations in the $\LLc$ interaction
and the second value reflects the uncertainty due to the employed regularization
scheme.

Our values are remarkably close to the ones of the $X(4660)$ resonance that have been
established in the reaction $e^+e^- \to \pi^+\pi^-\psi(2S)$ \cite{Lees:2014,Wang:2015}.
Therefore, we confirm a conjecture that has been already put forward shortly after
the $\eeLLc$ data were published, namely that the $X(4630)$ and $X(4660)$
resonances could be the same states \cite{Bugg:2009,Cotugno:2010,Guo:2010}.
We want to emphasize, however, that the present work takes into account the rather 
delicate interplay between the resonance and a possible residual interaction in 
the $\LLc$ system for the first time in a compelling way. Because of that we consider 
the outcome of the present analysis to be more conclusive. In particular, 
results could be achieved that are reliable on a quantitative level.

Finally, since new measurements for the reaction $\eeLLc$ are presently performed
by the BESIII Collaboration, with higher statistics and better energy
resolution \cite{BESIII}, we presented also predictions for the $\Lambda_c$ electromagnetic
form factors in the timelike region.  Indeed, our approach is well suited to perform also
calculations (and an analysis) of other and more subtle observables of the reaction $\eeLLc$
such as angular distributions, polarizations, or spin-correlation parameters,
once they become available \cite{BESIIIA}.

\section*{Acknowledgements}

We would like to thank Christoph Hanhart for useful comments and a careful reading 
of our manuscript. 
This work is supported in part by the DFG and the NSFC through
funds provided to the Sino-German CRC 110 ``Symmetries and
the Emergence of Structure in QCD''  (grant no. TRR 110)
and the BMBF (contract No. 05P2015 -NUSTAR R\&D).
The work of UGM was supported in part by The Chinese Academy
of Sciences (CAS) President's International Fellowship Initiative (PIFI) grant no.~2017VMA0025.


\end{document}